\documentclass[aps,pra,reprint,twocolumn,amsmath,superscriptaddress,amssymb,showpacs]{revtex4-1}
\usepackage{amsmath,braket,amssymb}
\usepackage[final]{graphicx}
\usepackage{subfigure}
\usepackage{xcolor}
\usepackage[english]{babel}
\usepackage[utf8]{inputenc}
\usepackage{color}
\usepackage{mathtools}
\usepackage{tikz-cd} 
\usepackage{tikz}

\usepackage[colorlinks,citecolor=darkBlue,linkcolor=darkBlue,
urlcolor=blue,hyperindex]{hyperref}

\usepackage[normalem]{ulem}
\normalfont

\definecolor{forestgreen}{rgb}{0.08, 0.4, 0.13}
\definecolor{darkBlue}{rgb}{0.08, 0.13, 0.4}

\begin{document}

\title{Enhanced entanglement negativity in boundary driven monitored fermionic chains}

\author{Xhek Turkeshi}
\affiliation{JEIP, USR 3573 CNRS, Coll\`{e}ge de France, PSL Research University, 11 Place Marcelin Berthelot, 75321 Paris, France}
\author{Lorenzo Piroli}
\affiliation{Philippe Meyer Institute, Physics Department, Ecole Normale Superieure, 
Université PSL, 24 rue Lhomond, 75231 Paris, France}
\author{Marco Schir\'o}
\affiliation{JEIP, USR 3573 CNRS, Coll\`{e}ge de France, PSL Research University, 11 Place Marcelin Berthelot, 75321 Paris, France}

\begin{abstract}
We investigate entanglement dynamics in continuously monitored open quantum systems featuring current-carrying non-equilibrium states. We focus on a prototypical one-dimensional model of boundary-driven non-interacting fermions with monitoring of the local density, whose average Lindblad dynamics features a well-studied ballistic to diffusive crossover in transport. Here we analyze the dynamics of the fermionic negativity, mutual information, and purity along different quantum trajectories. 
We show that monitoring this boundary-driven system enhances its entanglement negativity at long times, which otherwise decays to zero in absence of measurements. This result is in contrast with the case of unitary evolution where monitoring suppresses entanglement production. For small values of $\gamma$, the stationary-state negativity shows a logarithmic scaling with system size, transitioning to an area-law scaling as $\gamma$ is increased beyond a critical value. Similar critical behavior is found in the mutual information, while the late-time purity shows no apparent signature of a transition, being $O(1)$ for all values of $\gamma$. Our work unveils the double role of weak monitoring in current-driven open quantum systems, simultaneously damping transport and enhancing entanglement.
\end{abstract}

\maketitle

\section{Introduction}

Understanding entanglement dynamics in many-body systems is a fundamental challenge that bridges condensed matter to quantum information theory and helps us to characterize the wide spectrum of possible dynamical phases of matter. 
While generic isolated systems with local interactions exhibit universal features of  entanglement growth~\cite{zhou2020entanglement,nahum2017quantum},  different scenarios for entanglement behavior arise in the presence of ergodicity breaking, for instance due to many-body localization~\cite{abanin2019colloquium}, kinetic constraints~\cite{pichler2016realtime,surace2020lattice,desaules2021quantum,serbyn2021quantum}, long-range interactions~
\cite{schachenmayer2013entanglement,pappalardi2017multipartite,pappalardi2018scrambling,pappalardi2019entanglement,pappalardi2020origin,lerose2020bridging} or integrability~\cite{calabrese2005evolution,fagotti2008evolution,alba2017entanglement,alba2018entanglement,alba2019entanglement,alba2019quantum,alba2019quantum,alba2021generalized}. 


Entanglement is commonly believed to be destroyed by bulk coupling to a noisy environment~\cite{breuer2002theory}. This is usually the case for open-system dynamics which are described by a Lindbladian master equation and in which the environment is dealt with as a black box. A different behavior emerges by considering an open-system dynamics induced by \textit{monitoring} the system~\cite{plenio1998the, Wiseman2009}: in this setting, the environment is given by the measurement apparatus, which allows us to gain information on its state. The renewed interest in this type of dynamics has been largely motivated by the progress in quantum optics experiments, allowing us to manipulate and probe quantum systems to an unprecedented degree of control~\cite{basche1995direct,gleyzes2007quantum,vijay2011observation,robledo2011high,minev2019to}.

The possibility of monitoring the many-body dynamics has already proven its potential to realize new non-equilibrium phases, as explicitly shown in the simplest case where local unitary evolution is interspersed by local measurements~\cite{li2018quantum,li2019measurementdriven,chan2019unitary,skinner2019measurementinduced,nahum2021measurement,potter2021entanglement,barratt2021field,noel2021observation}. 
In this setting, extensive theoretical research~\cite{gullans2020dynamical,gullans2020scalable,bao2020theory,choi2020quantum,szyniszewski2019entanglement,snizhko2020quantum,turkeshi2020measurementinduced,sierant2022universal,sharma2022measurementinduced,ippoliti2021entanglement,klocke2022topological,ippoliti2021postselectionfree,ippoliti2022fractal,szyniszewski2020universality,kumar2020quantum,lunt2021measurementinduced,lunt2021quantum,li2021statistical,medina2021entanglement,agrawal2021entanglement,zabalo2020critical,zabalo2022operator,jian2019measurementinduced,lu2021spacetime,sharma2022measurementinduced,turkeshi2020measurementinduced,sierant2022universal,sierant2022dissipativefloquet,turkeshi2021measurementinduced_0} has provided strong evidence that generic systems described by random unitary circuits undergo a new type of measurement induced phase transition (MIPT), characterized by a change in the scaling of the subsystems entanglement entropy, from volume-law to area-law. On the other hand, certain classes of non-interacting systems under different types of monitoring protocols have been shown to display MIPT between a phase with sub-extensive entanglement growth and an area law~\cite{alberton2021entanglement,turkeshi2021measurementinduced,buchold2021effective,turkeshi2021entanglement,muller2022measurement,kells2021topological,turkeshi2022entanglementand}.

In this work, we put forward a different setting where monitoring might be expected to give rise to novel nonequilibrium entanglement behavior. Namely, we focus on current-driven open many-body quantum systems featuring non-thermal non-equilibrium stationary states (NESS). For concreteness, we consider the case of a system with a local $\mathrm{U}(1)$ charge, coupled at its ends to two reservoirs at different chemical potentials -- a prototypical framework to investigate quantum transport~\cite{bertini2021finitetemperature, landi2021non}. Accordingly, the dynamics under study is characterized by three ingredients: unitary evolution, boundary driving and monitoring of the $\mathrm{U}(1)$ charge. 

Current-driven setups are known to give rise to a non-trivial interplay between transport and entanglement. For instance, current-carrying states of non-interacting diffusive fermions can sustain extensive entanglement~\cite{gullans2019entanglement,gullans2019localization}, in stark contrast to the universal area-law scaling characterizing thermal phases~\cite{wolf2008area,sherman2016nonzero}. 
In the presence of monitoring, one may ask how trajectory-resolved features of entanglement depend on its rate. This question is particularly natural in light of the apparent competition between different effects: on the one hand, the driving forces a particle flow through the system while, on the other, large monitoring tends to pin it to an eigenstate of particle-density operators. 

\begin{figure}
	\centering
    \includegraphics[width=\columnwidth]{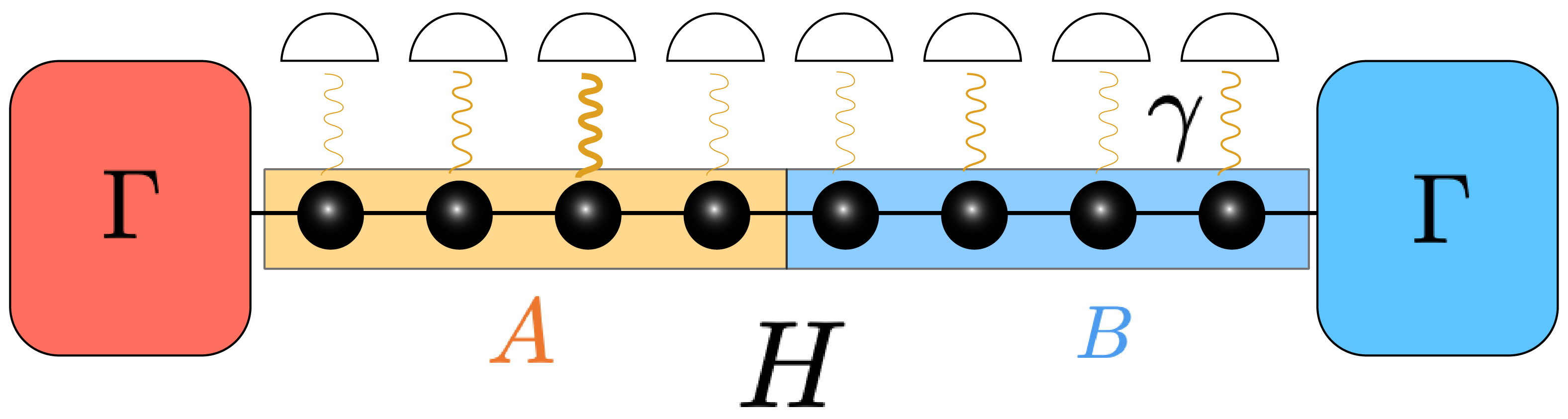}
	\caption{\textit{Pictorial representation of the setup. ---} A chain of fermionic degrees of freedom is subject to unitary dynamics dictated by the Hamiltonian $H$, weak monitoring of strength $\gamma$, and particle injection/depletion of the left/right edge sites. The negativity is computed between the two halves of the system $A$ and $B$. }
	\label{fig:scheme}
\end{figure}

We will address such a question by studying a simple but prototypical model of quantum transport: a one-dimensional chain of non-interacting fermions, where particles are injected and extracted at the two ends, respectively. In addition we will consider continuously monitoring the fermionic particle number, according to the so-called quantum-state-diffusion (QSD) protocol~\cite{caves1987quantum,diosi1998non,gisin1992quantum}, (see Fig~\ref{fig:scheme}). 

This model was introduced in Ref.~\cite{bernard2018transport} where its transport properties were analyzed within the limit of infinite monitoring rate. We also note that, in the absence of boundary driving terms, it coincides with that studied in Refs.~\cite{cao2019entanglement,alberton2021entanglement,coppola2022growth,buchold2021effective}. In Ref.~\cite{alberton2021entanglement,buchold2021effective}, in particular, it was shown that the system undergoes a MIPT between two phases, with logarithmic and area-law scaling of the entanglement entropy~\footnote{The absence of a volume-law phase is consistent with the non-interacting nature of the model, limiting its ability to scramble quantum information~\cite{fidkowski2021how,choi2020quantum} }.  While our analysis builds upon these works, the boundary driving introduces crucial differences. Most prominently, because the particle reservoirs are treated as an inaccessible environment, the state of the system along each quantum trajectory is a mixed state. Accordingly, in order to quantify the corresponding entanglement, we need to rely on the so-called negativity~\cite{eisert1999comparison,vidal2022computable,plenio2005logarithmic,shapourian2017partial,shapourian2019entanglement,shapourian2019twisted,murciano2021symmetry,murciano2022negativity}, since the entanglement entropy is known not to be a genuine measure of entanglement for mixed states~\cite{nielsen2002quantum}. 

In our model, the evolution of the average density matrix is described by a well studied Lindbladian master equation, with boundary driving and bulk dephasing~\cite{znidarivc2010exact, eisler2011crossover,znidaric2014exact,znidaric2014large,carollo2017fluctuating,turkeshi2021diffusion}. At late time, any non-zero rate of monitoring leads to diffusive behavior, thus damping the ballistic transport characterizing the isolated system. On the contrary, we show that monitoring can \emph{enhance} entanglement. Specifically, we find that the negativity displays a logarithmic growth for small values of $\gamma$, finally transitioning to an area-law scaling as $\gamma$ is increased beyond a critical value. 

Before leaving this section, we note that two recent works also explored the role of a dissipative environment for a system that is simultaneously monitored~\cite{weinstein2022measurement,ladewig2022monitored}, although from a different point of view. Ref.~\cite{ladewig2022monitored} considered a fermionic chain with a coupling to a dephasing bath at each bulk site in the system.  Ref.~\cite{weinstein2022measurement}, instead, focused on quantum circuit models with boundary dephasing, but featuring no local conservation law and therefore no current driving.

The rest of this paper is organized as follows. In Sec.~\ref{sec:fermions}  we introduce the model and the protocol we consider. In Sec.~\ref{sec:negativity} we study the purity and entanglement in the late-time regime, providing evidence of a transition in the scaling of the negativity. Our conclusions are consigned to Sec.~\ref{sec:conclusion}. Finally, the most technical aspects of our work are reported in several appendices.

\section{Monitored Driven Fermions}
\label{sec:fermions}

We begin by describing in detail the model studied in this work (cf. Fig.~\ref{fig:scheme}). We consider a one-dimensional chain of spinless fermions, governed by the Hamiltonian
\begin{equation}\label{eq:hamiltonian}
	{H}=-\sum_{j=1}^{L-1}\left[{c}^\dagger_j {c}_{j+1}+{c}^\dagger_{j+1} {c}_{j}\right]\,,
\end{equation}
where $L$ is the system size, while ${c}_j$, ${c}^\dagger_j$ are canonical fermionic operators. The system is coupled at its boundaries to particle reservoirs at different chemical potentials, and subject to continuous monitoring in the bulk. Within the so-called QSD protocol~\cite{caves1987quantum,diosi1998non,gisin1992quantum}, the evolution of the system density matrix $\rho$ is captured by the stochastic master equation (SME)
\begin{align}
	d{\rho}_\xi &= dt \mathcal{L}[{\rho}_\xi] + \sum_{m=1}^L d\Xi_m[{\rho}],\label{eq:noise}\\
	\mathcal{L}[\circ] &=-i [{H},\circ]  + \mathcal{D}_\mathrm{bnd}[\circ] + \mathcal{D}_\mathrm{bulk}[\circ].\label{eq:lindbladian}
\end{align}
The first term in Eq.~\eqref{eq:noise} describes the deterministic part of the evolution and includes several contributions, explicitly reported in Eq.~\eqref{eq:lindbladian}. The first one is a coherent term encoding the bulk unitary dynamics driven by the Hamiltonian~\eqref{eq:hamiltonian}. The second one corresponds to the boundary driving and reads $\mathcal{D}_\mathrm{bnd}[\circ]=\mathcal{D}^{L}_\mathrm{bnd}[\circ]+\mathcal{D}^{R}_\mathrm{bnd}[\circ]$, with
\begin{align}\label{eq:boundary_driving}
	\mathcal{D}^L_\mathrm{bnd}[\circ] &=\Gamma_L\left[\frac{1+\mu}{2}\left(2{c}^\dagger_1 \circ {c}_1 - \{{c}_1 {c}^\dagger_1 ,\circ\}\right)\right.\nonumber\\
	&\qquad \left.+\frac{1-\mu}{2}\left( 2{c}_1 \circ {c}^\dagger_1 -\{ {c}^\dagger_1 {c}_1 ,\circ\}\right)\right]\,,
\end{align} 
and
\begin{align}\label{eq:boundary_driving_2}
	\mathcal{D}^R_\mathrm{bnd}[\circ] &=\Gamma_R\left[\frac{1-\mu}{2}\left(2{c}^\dagger_L \circ {c}_L - \{{c}_L {c}^\dagger_L ,\circ\}\right)\right.\nonumber\\
	&\qquad \left.+\frac{1+\mu}{2}\left( 2{c}_L \circ {c}^\dagger_L -\{ {c}^\dagger_L {c}_L ,\circ\}\right)\right]\,.
\end{align} 
These two terms describe injection/depletion of particles at the two edges of the chain, respectively with rates 
$\Gamma_L(1\pm \mu)/2$ and $\Gamma_R(1\mp \mu)/2$. In the following we consider equal overall scales $\Gamma_L=\Gamma_R=\Gamma$ and unless stated otherwise set $\mu=1$, corresponding to the maximum difference in driving potential i.e. pure injection/depletion on the left/right edge. We will discuss the dependence on $\mu$ in Sec.~\ref{sec:purity}. Finally, the last term in~\eqref{eq:lindbladian} encodes the deterministic back-action due to bulk monitoring of the local particle density ${n}_i= {c}^\dagger_i {c}_{i}$, which takes the form of a dephasing dissipator
\begin{equation}
\mathcal{D}_\mathrm{bulk}[\circ] = -\frac{\gamma}{2}\sum_{j=1}^L [{n}_i,[{n}_i,\circ]]\,.
\label{eq:bulkdep}
\end{equation}
Monitoring is also responsible for the stochastic feedback term in ~\eqref{eq:noise}, which is defined as
\begin{equation}
d\Xi_m[\circ] \equiv \sqrt{\gamma} d\xi_t^m \{ {n}_m-\langle {n}_m\rangle_t^\xi,\circ\}\,.\label{eq:backfeed}
\end{equation}
It is characterized by independent Brownian processes, with \^Ito differentials  $d\xi^i_t$ such that $d\xi^i_t d\xi^j_{t'} = \delta(t-t') \delta_{i,j} dt$. In Eq.~\eqref{eq:backfeed} we introduced the notation $\langle {A}\rangle_t^\xi\equiv \mathrm{tr}({\rho}_\xi {A})$. 
We emphasize the difference of our setting with respect to Ref.~\cite{ladewig2022monitored}, where an additional bulk dephasing channel was added on top of the monitoring process.

The solution to the SME~\eqref{eq:lindbladian} is a conditional density matrix ${\rho}_\xi$ which encodes complete information on the monitored quantum system. For a given functional of the density matrix, $\mathcal{F}[\cdot]$, we may define its statistical distribution as
\begin{equation}
	P(F) \equiv \int [d\xi]P(\xi)\delta(\mathcal{F}[\rho_\xi]-F).
\end{equation}
A natural class of functionals is the expectation value of an observable ${O}$, that is $\mathcal{F}_{{O}}[\rho]\equiv \mathrm{tr}(\rho {O})$. In this case, the average over the trajectories 
\begin{equation}
	\overline{O}\equiv \int dO P(O) O,
\end{equation}
coincides with the trace over the average density matrix $\mathrm{tr}({O}\overline{\rho})$, where 
\begin{equation}
	\overline{\rho} = \int [d\xi] P(\xi) \rho_\xi\,.
\end{equation}
It is easy to show that $\overline{\rho}$ satisfies
\begin{equation}\label{eq:averaged_dynamics}
	\frac{{\rm d}}{{\rm d}t}\bar{\rho}=\mathcal{L}[\bar{\rho}]\,.
\end{equation} 
This Lindbladian equation has been extensively studied in the literature~\cite{znidarivc2010exact, eisler2011crossover,znidaric2014exact,znidaric2014large,carollo2017fluctuating,turkeshi2021diffusion}, and the structures of its late-time NESS has been worked out analytically~\cite{znidarivc2010exact}. For completeness, we review its main properties in Appendix~\ref{sec:NESS}.

When considering non-linear functionals of the density matrix, the average behavior can not be computed from the averaged density matrix. This is the case, in particular, for the purity and the entanglement negativity, discussed in Sec.~\ref{sec:negativity}. In general, in order to study the dynamics of these quantities, one needs to solve the SME and sample over different trajectories. 

Crucially, in the non-interacting model~\eqref{eq:noise}, the dynamics along each trajectory can be computed efficiently starting from a Gaussian initial state~\cite{bravyi2005lagrangian}. Indeed, a fermionic Gaussian state evolved under the SME~\eqref{eq:boundary_driving} remains Gaussian ~\cite{bravyi2005lagrangian}, so that, along each given quantum trajectory, the state of the system is completely characterized by its \emph{covariance matrix}. 

If the initial state is such that 
\begin{equation}
\label{eq:condition}
\langle {c}^\dagger_i {c}^\dagger_j\rangle_{t=0} =0\,,
\end{equation}
 it is easy to see that this remains true at later times, and the state of the system is completely encoded into the matrix
\begin{equation}
	(C^t_\xi)_{m,n}\equiv \langle {c}^\dagger_m {c}_n\rangle_t^\xi.\label{eq:corrmat}
\end{equation}
Accordingly, the full many-body evolution, that is a trajectory in the $2^{2L}$-dimensional space of fermionic density matrices, can be replaced by the evolution of the $L\times L$ covariance matrix~\eqref{eq:corrmat}. The explicit stochastic equation satisfied by the covariance matrix is reported in Appendix~\ref{sec:derivation_master}, where we also provide details on our numerical solution.

\section{Entanglement negativity and purification}
\label{sec:negativity}

In this section, we present our results for the dynamics of entanglement, as quantified by the so-called \emph{fermionic negativity}. We will also discuss the behavior of mutual information and study the purity of the system.

\subsection{Fermionic negativity}

As already mentioned, because the state of the system along each quantum trajectory is mixed, the von Neumann entropy is not a good measure of quantum entanglement~\cite{nielsen2002quantum}. Here, we focus on the fermionic negativity~\cite{shapourian2017partial}, which has been recently proposed as a measure of mixed-state entanglement alternative to the logarithmic negativity~\cite{eisert1999comparison,vidal2022computable,plenio2005logarithmic}. Contrary to the latter, the fermionic negativity can be computed efficiently for fermionic Gaussian states~\cite{shapourian2017partial}, while it is also a genuine entanglement monotone~\cite{shapourian2019entanglement}. In the context of Lindbladian dynamics of non-interacting fermionic chains, the fermionic negativity has been extensively investigated~\cite{alba2021spreading,alba2022logarithmic}, and its behavior has been understood based on a semi-classical quasi-particle picture. Here, we show that qualitative differences arise in the presence of monitoring. 

The fermionic negativity is defined based on the concept of partial time reversal~\cite{shapourian2017partial}, as we now briefly summarize. Let us introduce the Majorana operators $\psi_{2k-1} = {c}_k+{c}_k^\dagger$ and $\psi_{2k} = i ({c}_k-{c}_k^\dagger)$.  Given a bipartition $A\cup B$, and denoting by ${a}_j$ and ${b}_j$ the Majoranas acting respectively on $A$ and $B$, the system density matrix is given by
\begin{equation}
    {\rho} = \sum_{k_1,k_2}^{k_1+k_2\ \mathrm{even}} \rho_{p_1,\dots,p_{k_1}}^{q_1,\dots,q_{k_2}} \prod_{i=1}^{k_1} {a}_{p_i} \prod_{j=1}^{k_2} {b}_{p_j}.
\end{equation}

\begin{figure}
    \centering
    \includegraphics[width=\columnwidth]{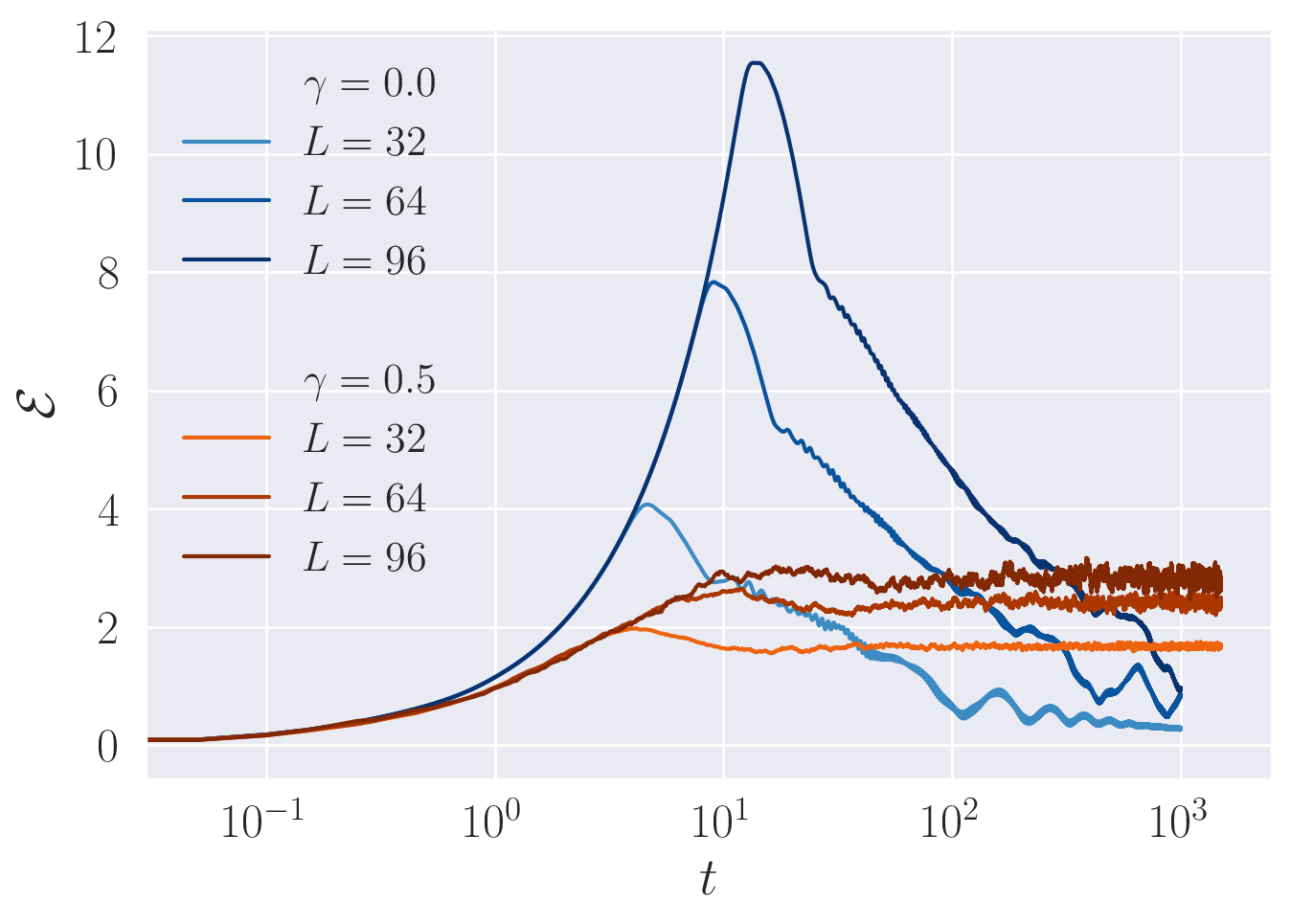}
    \caption{\label{fig:negcomp} Entanglement negativity as a function of time. We plot two sets of curves corresponding to zero ($\gamma=0.0$) and weak ($\gamma=0.5$) monitoring. In the two cases, we consider chains of size $L=32\div 96$ with $\Gamma=1$ and $\mu=1$. 
    At late times, the negativity vanishes in non-monitored systems, while it approaches a constant increasing with $L$ in the presence of weak monitoring. }
\end{figure}

\noindent Introducing the partial time reversal on the subsystem $A$ 
\begin{equation}
    {\rho}^{R_A} = \sum_{k_1,k_2}^{k_1+k_2\ \mathrm{even}} \rho_{p_1,\dots,p_{k_1}}^{q_1,\dots,q_{k_2}} i^{k_1} \prod_{i=1}^{k_1} {a}_{p_i} \prod_{j=1}^{k_2} {b}_{p_j},\label{eq:rhoR1}
\end{equation}
the fermionic negativity is defined as
\begin{equation}
    \mathcal{E}(\rho;A) = \ln \left[\mathrm{tr}\sqrt{{\rho}^{R_A} ({\rho}^{R_A})^\dagger}\right]\,.\label{eq:defneg}
\end{equation}
It is a measure of the entanglement of the state $\rho$ shared between the regions $A$ and $B$. Importantly, if the state $\rho$ is Gaussian, so is $\rho^{R_A}$ and, in this case, the entanglement negativity can be obtained with polynomial computational resources. For completeness, we detail the procedure to compute it in Appendix~\ref{sec:fermionic_neg}, while here we only report the final results of our analysis.

We focus on the partition given in Fig.~\ref{fig:scheme}, with $|A|=L/2$ and begin with the real-time evolution of the average negativity
\begin{equation}
\mathcal{E}(t)\equiv\mathbb{E}_\xi [\mathcal{E}({\rho}_\xi; A)]\,,
\end{equation}
from the initial state
\begin{equation}
\label{eq:neel_state}
\ket{\Psi_0}=c^{\dagger}_1 c^\dagger_3 \cdots c^\dagger_{L-1}\ket{0}\,,
\end{equation}
where $\ket{0}$ is the vacuum. $\ket{\Psi_0}$ has no entanglement, is Gaussian and satisfies~\eqref{eq:condition}. Therefore, we can apply the numerical scheme described in Appendix~\ref{sec:derivation_master}. We note that the late-time stationary state does not depend on the choice of the initial state, as we explicitly verified and discuss in Appendix~\ref{sec:additional_checks}; there we also detail the parameters used for the numerical simulations. 

An example of our numerical data is shown in Fig.~\ref{fig:negcomp}. In the absence of monitoring, we find that the negativity grows linearly in time, displaying an ``entanglement barrier''. Namely, after a time proportional to $L$, $\mathcal{E}(t)$ reaches a maximum value increasing linearly with the system size $L$. Within this time frame, we verified that there is a ballistic data collapse according to the scaling function $\varepsilon(t)=\mathcal{E}(t/L,\ell/L)/L$. At later times, the fermionic negativity starts decaying, as the quantum correlations are washed out by the boundary coupling with the dissipative environment. Our numerical data are compatible with a zero negativity in the infinite-time limit, for all system sizes.

The evolution is qualitatively different for $\gamma>0$. For weak monitoring, corresponding to $\gamma=0.5$ in Fig.~\ref{fig:negcomp}, the entanglement negativity $\mathcal{E}$ grows logarithmically in time, saturating to a non-zero value. Contrary to the case $\gamma=0$, the fermionic negativity does not show an entanglement barrier: this is in line with the expectation that monitored non-interacting fermionic dynamics can not sustain extensive entanglement, due to the absence of scrambling~\cite{cao2019entanglement,alberton2021entanglement,fidkowski2021how,choi2020quantum}. 
On the other hand, it is evident from Fig.~\ref{fig:negcomp} that, at late times, the negativity for $\gamma>0$ is larger than in the case $\gamma=0$: namely, monitoring a boundary-driven system causes an enhancement of steady-state entanglement, in stark contrast to what happens in monitored isolated-system evolution~\cite{li2018quantum,li2019measurementdriven,chan2019unitary,skinner2019measurementinduced,nahum2021measurement,potter2021entanglement,barratt2021field} where entanglement production is suppressed by measurement, with respect to the unitary dynamics.

\begin{figure}
    \centering
    \includegraphics[width=\columnwidth]{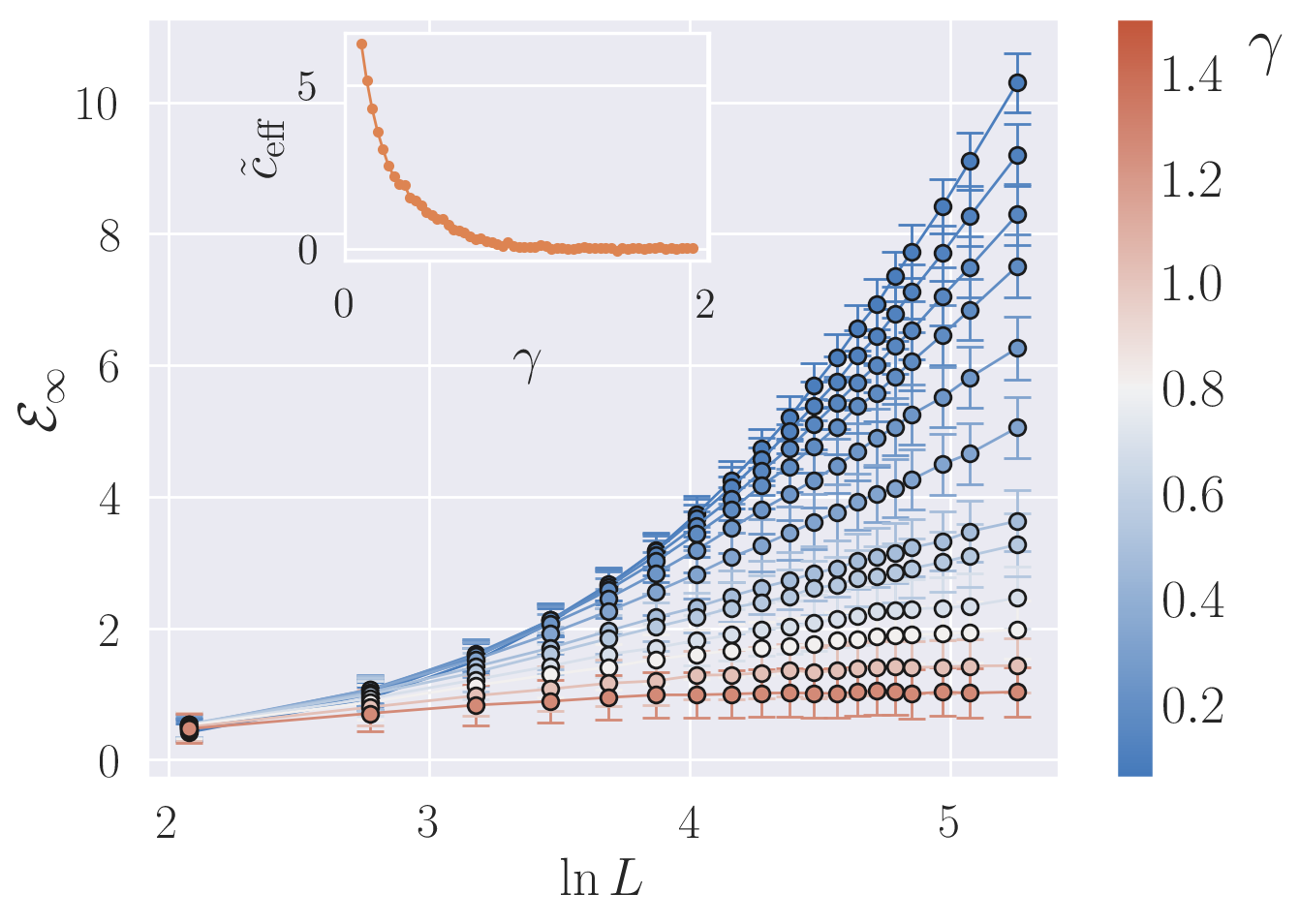}
    \caption{\label{fig:negativity} Large-time limit of the fermionic negativity. We consider chains of size $L= 8\div 192$ with $\Gamma=1$, $\mu=1$, and $\gamma=0.15\div 2 $.  Inset: numerical fit for the effective central charge appearing in Eq.~\eqref{eq:CFT_formula}. }
\end{figure}

As expected, we find that the late-time negativity depends on the monitoring strength. For small $\gamma$, we see that the scaling of the latter is consistent with a logarithmic growth in $L$. Instead, for $\gamma$ sufficiently large, the late-time stationary value of the negativity appears to be convincingly independent of $L$, despite being generically non-zero. In order to be quantitative, we analyze the system-size scaling of the stationary negativity $\mathcal{E}_\infty$ as a function of $L$, for $L\le 192$. Our results are reported in Fig.~\ref{fig:negativity}, from which we see evidence of a transition at a critical value, $\gamma=\gamma_c$ separating a logarithmic from an area-law scaling. In the small-$\gamma$ regime, we have performed a fit of the negativity against the formula
\begin{equation}\label{eq:CFT_formula}
    \mathcal{E} = \frac{\tilde{c}_\mathrm{eff}(\gamma)}{3}\ln L + e_0(\gamma)\,,
\end{equation}
obtaining an estimate for the effective central charge $\tilde{c}_\mathrm{eff}(\gamma)$~\cite{calabrese2004entanglement} and the constant $e_0(\gamma)$. Both parameters are found to continuously vary with $\gamma$, as shown in inset of Fig.~\ref{fig:negativity}. For $\gamma\gtrsim 0.8$, our fitting procedure gives us $\tilde{c}_\mathrm{eff}(\gamma)=0$, which allows us to identify $\gamma_c\simeq 0.8$ as the critical value separating the two phases. 

We have verified that our estimate for $\gamma_c$ does not depend on the strength of the boundary coupling $\Gamma$. In addition, it is consistent with the critical value found in Ref.~\cite{alberton2021entanglement} characterizing the MIPT in the isolated non-interacting fermionic chain (corresponding to $\Gamma=0$). Therefore,  the critical behavior of the bipartite entanglement in the stationary state appears to be dominated by the physics in the bulk, i.e. by the competition between unitary hopping and monitoring. We note however that boundary driving can influence bulk properties such as the density profile, cf. Appendix~\ref{sec:NESS}. 

We expect that this transition in the entanglement negativity can be understood based on an approach similar to the one developed in Ref.~\cite{buchold2021effective}, which introduced an effective $n$-replica Keldysh field theory capturing the average of the $n$-th moment of the quantum trajectories, followed by a bosonization of the bulk monitored problem. In this framework the boundary driving should appear as a local non-linearity for the effective replica field theory, not modifying the nature of the bulk transition. In turn, this would imply that the fermionic negativity undergoes a Berezinskii-Kosterlitz-Thouless (BKT) transition, as established in Refs.~\cite{alberton2021entanglement,buchold2021effective} for the entanglement entropy of the isolated system.

In order to substantiate further this claim, it would be useful to provide numerical results for larger system sizes, especially given the expected large finite-size effects characterizing the BKT transition. Unfortunately, we are not able to simulate systems of the same sizes studied in Ref.~\cite{alberton2021entanglement}. In our setting, the main limitation comes from the fact that we need to follow the evolution up to very large times, in order to reach the stationary regime. This can be understood at the level of the average Lindbladian dynamics, cf. Appendix~\ref{sec:NESS}. Indeed, because the latter displays diffusive behavior ~\cite{znidarivc2010exact}, one has that the stationary regime is approached at times $t\sim L^2/D$, where $D=(\gamma+1/\gamma)$ is the diffusion constant~\cite{eisler2011crossover}. Therefore, in our simulations we need to follow the dynamics up to times that scale quadratically in $L$, limiting the system sizes which can be analyzed.

\begin{figure}[t!]
	\centering
	\includegraphics[width=\columnwidth]{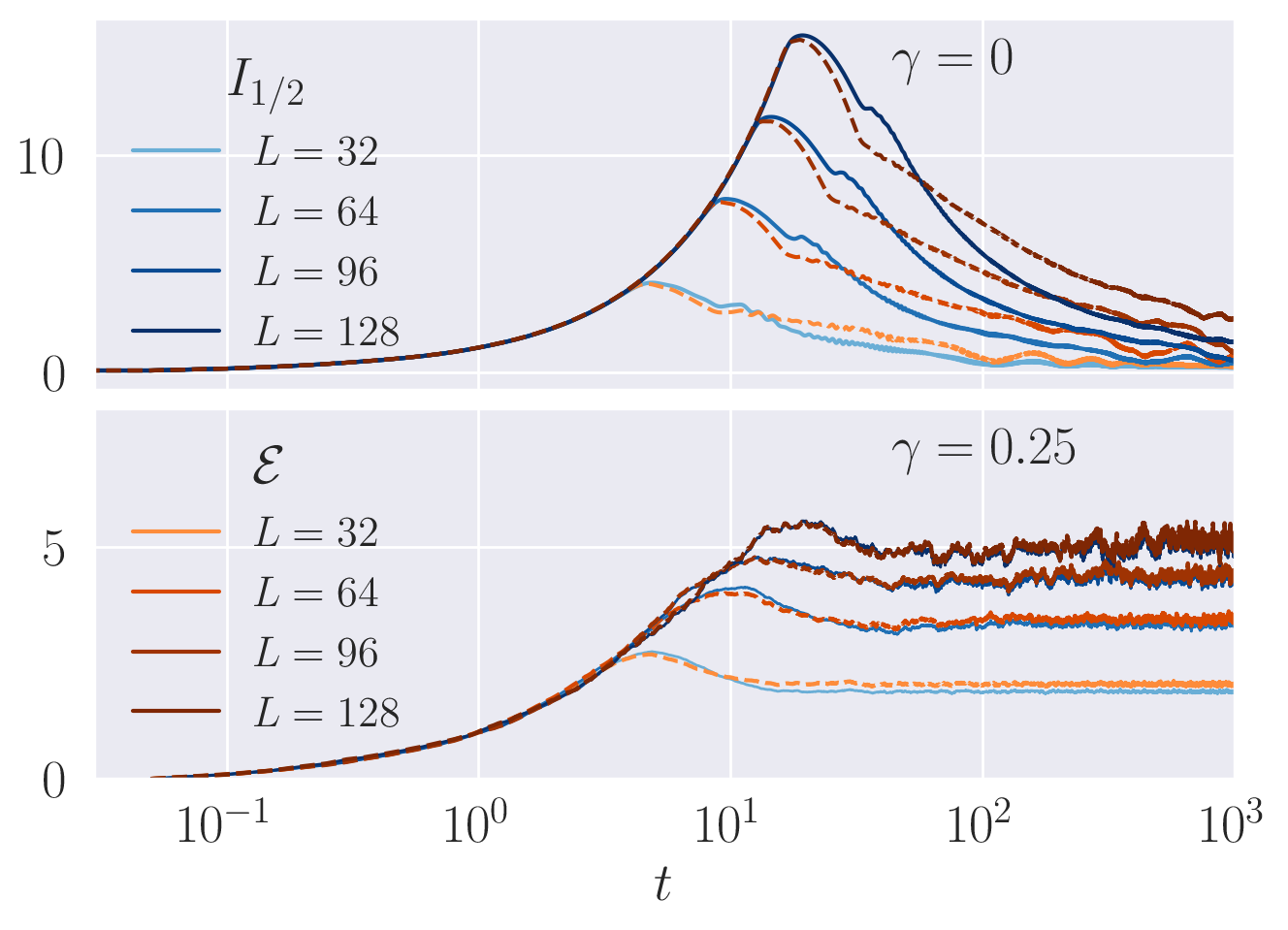}
	\caption{\label{fig:mutual} Comparison between the dynamics of R\'enyi-$1/2$ mutual information and fermionic negativity, from the initial state~\eqref{eq:neel_state}. We consider chains of size $L= 32\div 128$ with $\Gamma=1$, $\mu=1$, and $\gamma=0.0$, $\gamma=0.25$. }
\end{figure}
Together with the fermionic negativity, we have also studied the R\'enyi mutual information associated with the bipartition displayed in Fig.~\ref{fig:scheme}. It is defined as
\begin{equation}
I_{A: B}^{(\alpha)}(t):=S_{A}^{(\alpha)}(t)+S_{B}^{(\alpha)}(t)-S_{A B}^{(\alpha)}(t)\,,
\end{equation}
where
\begin{align}
	S_{S}^{(\alpha)}(t)&=\mathbb{E}_\xi \left[S_{S}[{\rho_\xi}]\right] \\
	S_{S}[{\rho_\xi}]&=\frac{1}{1-\alpha} [\ln \operatorname{tr}\left[\rho^{S}_\xi(t)^{\alpha}\right]]\,,
\end{align}
is the R\'enyi entropy and $\rho^{S}_\xi(t)$ is the density matrix reduced to the subsystem $S$. 
It is a measure of both classical and quantum correlations~\cite{wolf2008area} which can be non-zero even for non-entangled states.  We recall that the R\'enyi entropy of a subsystem for a Gaussian state can be efficiently calculated~\cite{vidal2003entanglement,Peschel_2009}. Given the reduced correlation matrix $C^S_{i,j} = C_{i,j}$ for $i,j\in S$, the R\'enyi entropy is given by
\begin{equation}
    S_{S}[{\rho_\xi}] = \frac{1}{1-\alpha}\mathrm{tr}\ln \left[ (C_\xi^S)^\alpha + (\openone-C^S_\xi)^\alpha\right]\,.~\label{eq:effent}
\end{equation}

For unitary quench dynamics in non-interacting fermionic chains, it has been shown~\cite{alba2019quantum} that, in a scaling limit of large system sizes and times, the fermionic negativity of a bipartition is proportional to the R\'enyi-$1/2$ mutual information, i.e.
\begin{equation}\label{eq:identification_i_e}
\mathcal{E}=\frac{1}{2} I_{A: B}^{(1 / 2)}\,.
\end{equation}
Recently, this identification has been extended to arbitrary unitary quantum-circuit dynamics up to times linear in the subsystem sizes~\cite{bertini2022entanglement}. On the other hand, Eq.~\eqref{eq:identification_i_e} does not generally hold for non-unitary evolution, as shown for non-interacting fermionic chains with dephasing noise~\cite{alba2022logarithmic}.

Motivated by these discussions, we have probed the validity of Eq.~\eqref{eq:identification_i_e} in the presence of monitoring. An example of our results is shown in Fig.~\ref{fig:mutual}. Interestingly, despite the dynamics being non-unitary, we found that Eq.~\eqref{eq:identification_i_e} is exactly verified up to times proportional to the system size, both for $\gamma=0$ and $\gamma>0$. At later times, the two deviate from one another, but still remain numerically close and display the same qualitative behavior. In particular, the mutual information shows the same phase transition of the fermionic negativity. We note that a similar relation between the two quantities was also found in Ref.~\cite{weinstein2022measurement} studying monitored quantum circuits with boundary dephasing.

\subsection{The purity}\label{sec:purity}

Finally, we study the dynamics of the purity of the total system, i.e.
\begin{equation}
\mathcal{P}(t)=\mathbb{E}_\xi[{\rm tr}[\rho_{\xi}^2]]\,.
\end{equation}
For a single quantum trajectory, it is related to the R\'enyi-$2$ entropy via $\mathcal{P}[\rho_\xi]=e^{-S^{(2)}[\rho_\xi]}$. This observation allows us to compute it efficiently, cf. Eq.~\eqref{eq:effent}.

For non-interacting fermionic systems, a mixed state subject to monitored unitary dynamics \emph{purifies}, i.e. becomes a pure state, in a time that is polynomial in the system size~\cite{fidkowski2021how}. However, this is in general not true in the presence of an incoherent coupling to the environment, since the latter tends to increase the system entropy, competing with the effect of monitoring. In our setting, a natural question then pertains to the scaling of the purity in the late-time stationary regime, and whether the latter is able to diagnose a transition as a function of $\gamma$. A similar question was also addressed recently in Ref.~\cite{ladewig2022monitored}.

\begin{figure}[t!]
	\centering
	\includegraphics[width=\columnwidth]{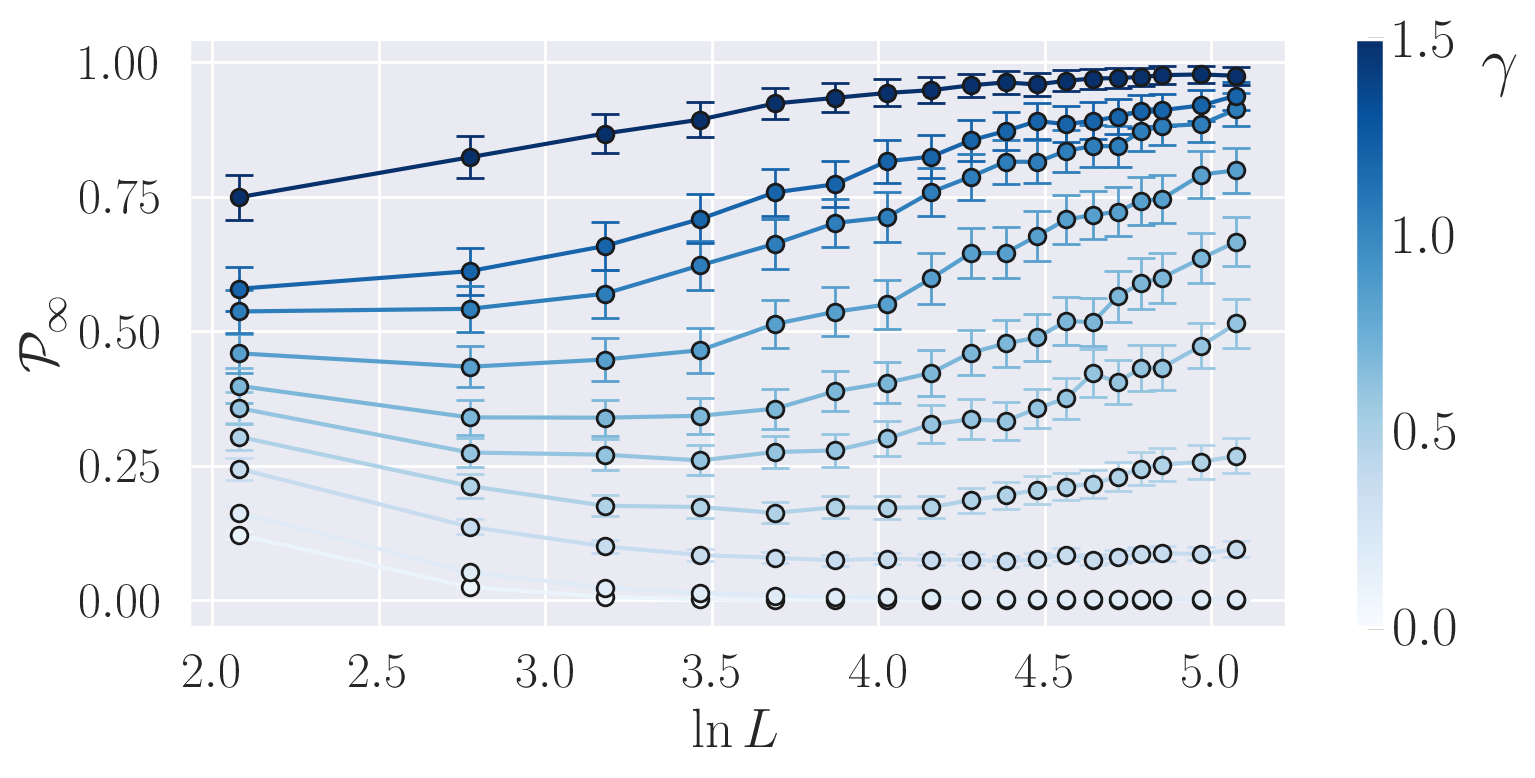}
	\caption{\label{fig:purity} Large-time limit of the averaged purity. We consider chains of size $L=8\div 192$ with $\Gamma=1$, $\mu=1$, $\gamma=0.15\div 1.5$.}
\end{figure}

We have found that, contrary to the fermionic negativity, the purity has a strong dependence on $\mu$, modeling the driving potential difference, cf. Eqs.~\eqref{eq:boundary_driving} and \eqref{eq:boundary_driving_2}. We begin by discussing our results in the simplest case $\mu=1$. Our numerical data for the late-time limit of the purity for different values of $\gamma$ are displayed in Fig.~\ref{fig:purity}. The plots indicate that the purity remains close to $1$ for large $L$, so that its logarithm is $O(1)$ as $L\to\infty$. For small $\gamma$, the scaling of $\mathcal{P}$ shows a non-monotonic behavior, with an asymptotic growth towards $1$. This trend is manifest well below the critical value $\gamma_c\sim 0.8$ and is visible, within the accessible system sizes, at least down to $\gamma\sim 0.3$. 

These numerical results suggest that the purity scaling does not display a transition as a function of $\gamma$, and in particular it is never vanishing as $L\to\infty$. Although we were not able to prove this rigorously, in Appendix~\ref{sec:asymptotic_purity} we provide a simple heuristic argument to justify it. In essence, the idea is that, as $L\to\infty$, the particle densities in the left-most and right-most sites are close to $1$ and $0$, respectively. This follows from the knowledge of the average density profile in the Lindbladian steady state, as discussed in Appendix~\ref{sec:NESS}. Accordingly, the injection and depletion of particles at the ends of the chain is suppressed, damping the rate of entropy growth due to the boundary Lindbladian. Our argument is completed by combining this picture with a lower bound on the purification rate for non-interacting fermionic systems~\cite{fidkowski2021how}.

\begin{figure}
	\centering
	\includegraphics[width=\columnwidth]{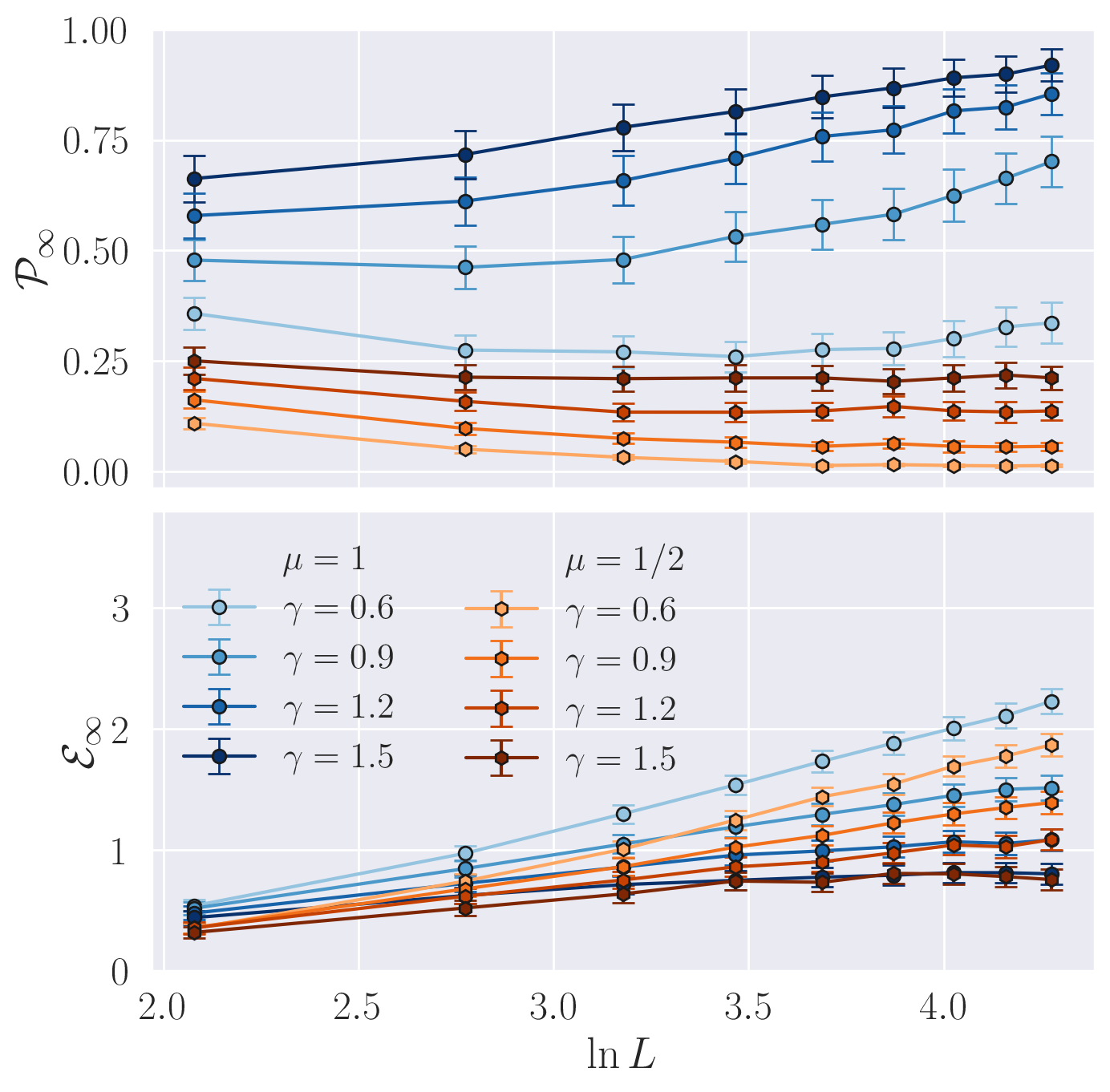}
	\caption{\label{fig:mu} Large-time limit of the averaged purity and entanglement negativity, for different choices of the driving potential difference $\mu$. We fix $\Gamma=1$ and $\gamma=0.6\div 1.5$. As detailed in the text, the purity is qualitatively sensible to the values of $\mu$, while the negativity is only quantitatively affected by $\mu$, the scaling being qualitatively the same.  }
\end{figure}

This discussion also suggests a non-trivial dependence of the purity from $\mu$, because it changes the particle densities at the boundary sites. We have verified this numerically, as shown in Fig.~\ref{fig:mu}, where we report data for the purity and negativity for different values of $\mu$ and $\gamma$. First, we see that the qualitative behavior of the negativity is independent from the boundary parameters, confirming the picture established in the previous section. On the other hand, the scaling of the purity is more complicated. For $\mu=0.5$ it appears to be vanishing for weak monitoring ($\gamma=0.6$), while remaining approximately constant for $\gamma\gtrsim 0.9$. 
Our numerics suggest a $\gamma$-dependent stationary value for $\mu<1$ with a finite stationary mixedness; however, the limited accessible system sizes are not conclusive to rule out that these are finite-size effects, and that for $\mu<1$ the purity vanishes for $L\to \infty$. 

\section{Conclusions}
\label{sec:conclusion}

We have investigated the entanglement dynamics in a prototypical one-dimensional model of boundary-driven non-interacting fermions in the presence of monitoring. We have shown in this context that the interplay between boundary dissipation and monitoring can enhance entanglement, as quantified by the fermionic negativity, as opposed to the unitary case where monitoring is detrimental to entanglement production. Furthermore, we have provided evidence that the system undergoes a phase transition that manifests itself in the scaling of the late-time entanglement negativity, going from a logarithmic to an area-law scaling. We have also shown that the transition can be diagnosed from the bipartite mutual information, but not from the purity of the whole system. Our results complement recent works studying the effect of a dephasing environment on monitored unitary dynamics~\cite{weinstein2022measurement,ladewig2022monitored}.

Our work raises several questions. First, it would be interesting to substantiate analytically our results on the MIPT of the entanglement negativity. As mentioned, we expect that a possible strategy could be to extend the field-theoretical approach developed in Ref.~\cite{buchold2021effective}. Second, a straightforward direction would be to investigate measurement dynamics in transport settings beyond the non-interacting case considered in this work. For instance, a natural question is how our findings are modified in the presence of additional unitary noise, such as in the Quantum Symmetric Simple Exclusion Process~\cite{bauer2017stochastic,bernard2019open,bauer2019equilibrium,jin2020stochastic,bernard2021entanglement,bernard2022dynamics,hruza2022dynamics}, or of interactions. We believe that an interesting tractable model to study the latter problem is given by random unitary circuits featuring a $U(1)$ conserved charge~\cite{khemani2018operator,rakovszky2018diffusive,piroli2020random}, with additional incoherent boundary terms implementation charge injection and extraction.

Perhaps, the most interesting direction pertains to the study of transport features beyond the average Lindbladian dynamics. In this respect, a non-trivial task is to define a meaningful notion of transport at the level of individual quantum trajectories, since the monitoring brings about violations of the local charge continuity equation. We leave these questions for future work.

\begin{acknowledgments}
LP acknowledges Denis Bernard for interesting discussions and collaborations on related topics. XT is grateful to V. Alba, P. Ruggiero, and V. Vitale for discussions. 
XT and MS were supported by the ANR grant “NonEQuMat”(ANR-19-CE47-0001). We acknowledge
computational resources on the Coll\'ege de France IPH cluster.
\end{acknowledgments}

\appendix 

\section{Equation of motion for the correlation matrix}
\label{sec:derivation_master}
In this section, we derive the equation of motion for the correlation matrix $C_\xi^t$. For readability, in this section, we drop the time label and the trajectory label.
Following the prescriptions in Refs.~\cite{cao2019entanglement,alberton2021entanglement,buchold2021effective,coppola2022growth,turkeshi2021measurementinduced}, the stochastic Schr\"odinger equation for the correlation matrix is easily derived and reads
\begin{widetext}
\begin{align}
    d C_{i,j} & = i dt (C_{i-1,j}+ C_{i+1,j}- C_{i,j+1}- C_{i,j-1}) -{\gamma} C_{i,j}dt    \nonumber \\ 
    &+ \gamma dt\sum_{m=1}^L C_{i,m} C_{m,j} + (1+\mu)\Gamma_\mathrm{L} \delta_{i,1}\delta_{j,1}dt + (1-\mu)\Gamma_\mathrm{R} \delta_{i,L}\delta_{j,L}dt \nonumber \\ 
  &\qquad - \left(\Gamma_\mathrm{L}\frac{(1+\mu)}{2}+\Gamma_\mathrm{R}\frac{(1-\mu)}{2}\right)(\delta_{i,1}+ \delta_{j,1}) C_{i,j}dt - \left(\Gamma_\mathrm{R}\frac{(1+\mu)}{2}+\Gamma_\mathrm{L}\frac{(1-\mu)}{2}\right)(\delta_{i,L}+ \delta_{j,L}) C_{i,j}dt \nonumber \\ &\qquad+(d\xi_{i,t}+d\xi_{j,t})C_{i,j} - 2 \sum_{m=1}^L C_{i,m} d\xi_{m,t} C_{m,j} .\label{eq:correlationeom1}
\end{align}
Up to a sub-leading Trotterization error, we can consider the equation of motion from two separate contributions: (i) the Hamiltonian and the boundary Lindbladian, (ii) the monitoring contribution. 

The former has been considered in a variety of works (see e.g. Refs.~\cite{znidarivc2010exact,esposito2005emergence,esposito2005exactly,turkeshi2021diffusion}), and is given by the terms in Eq.~\eqref{eq:correlationeom1} which are not proportional to $\gamma$ or to $d\xi_{j,t}$
\begin{equation}
    \frac{d}{dt} C = \mathbb{L}[C] + \mathbb{P}.\label{eq:schematic}
\end{equation}
The linear operator $\mathbb{L}$ and the matrix $\mathbb{P}$ are simply read out from the corresponding terms in Eq.~\eqref{eq:correlationeom1}, and Eq.~\eqref{eq:schematic} can be integrated with standard means (e.g. Runge-Kutta algorithms) to obtain the infinitesimal solution $\tilde{C}_{t+dt}$.

The noisy contribution can be integrated as well, to obtain~\cite{cao2019entanglement} 
\begin{equation}
    C_{t+dt} \propto e^{(d\vec{\xi}_t +\gamma dt (2C_\mathrm{diag}-\openone))})\tilde{C}_{t+dt}e^{(d\vec{\xi}_t +\gamma dt (2C_\mathrm{diag}-\openone))})\label{eq:zio}
\end{equation}
where $C_\mathrm{diag}$ is the diagonal part of $C$.
The transformation Eq.~\eqref{eq:zio} corresponds to
\begin{equation}
    \rho'  = e^{\sum_{m} d\xi^m_t (n_m-\langle n_m\rangle) - \gamma dt \sum_{m} (n_m-\langle n_m)^2}\rho e^{\sum_{m} d\xi^m_t (n_m-\langle n_m\rangle) - \gamma dt \sum_{m} (n_m-\langle n_m)^2}.
\end{equation}
We note that this transformation preserves the norm only up to terms $o(dt)$, and the errors accumulate during the evolution. Thus, we choose to renormalize the state after each step and consider instead
\begin{equation}
    \rho'  = \frac{e^{\sum_{m} d\xi^m_t (n_m-\langle n_m\rangle) - \gamma dt \sum_{m} (n_m-\langle n_m)^2}\rho e^{\sum_{m} d\xi^m_t (n_m-\langle n_m\rangle) - \gamma dt \sum_{m} (n_m-\langle n_m)^2}}{\mathrm{tr}\left[e^{2\sum_{m} d\xi^m_t (n_m-\langle n_m\rangle) - 2\gamma dt \sum_{m} (n_m-\langle n_m)^2}\rho\right]}.
\end{equation}
We can think of this map as a sequence of single-site commuting transformations. The associated map for the correlation matrix can be obtained within the framework of linear fermionic optics~\cite{bravyi2005lagrangian,fidkowski2021how}. The final result is given by the combined action of $L$ commuting channels $C(t+dt) = \mathcal{M}_1 \circ \mathcal{M}_2\circ\cdots\circ \mathcal{M}_L [\tilde{C}_{t+dt}]$, each of which is associated to a given site, and reads

\begin{equation}
    \mathcal{M}_j(C) = D^{(j)}\left[C+x_j (E^{j} C + C E^{(j)} - 2 C E^{(j)} C) - \frac{x_j+1}{2}E^{(j)}\right]D^{(j)} + \frac{\tanh(\varepsilon_j) + 1}{2}E^{(j)}\label{eq:finalff}
\end{equation}

where 
\begin{align}
    [D^{(j)}]_{m,n} &= \delta_{m,n} \left[\frac{1}{\cosh(\varepsilon_j)}\delta_{n,j} + (1-\delta_{n,j})\right],\\
    [E^{(j)}]_{m,n} &= \delta_{n,j} \delta_{m,j},\\
    x_{j} &= \frac{\tanh(\varepsilon_j)}{1-(1-2 C_{jj})\tanh(\varepsilon_j)}\\
    \varepsilon_j &= d\xi_t^j + (2 C_{jj} -1) dt.\label{eq:vareps}
\end{align}

\end{widetext}
\section{Average Lindbladian equation and NESS}
\label{sec:NESS}
In this section we give a brief summary of known results for the average stationary state $\overline{\rho} = \mathbb{E}_\xi[\rho_\xi]$. 
The density, current density and two-point density functions in the non-equilibrium steady state were computed in Ref.~\cite{znidarivc2010exact}.
The average current $j_\mathrm{NESS} = i \langle ({c}^\dagger_m {c}_{m+1} -{c}^\dagger_{m+1} {c}_{m} )\rangle_\mathrm{NESS}$ is given by
\begin{equation}
    j_\mathrm{NESS} =  -\frac{\mu}{\Gamma + \Gamma^{-1} + (L-1)\gamma/2}\label{eq:diffusive},
\end{equation}
and displays in the large system size limit $L\gg 1$ a diffusive behavior. 
The average density, on the other hand, shows a linearly decreasing gradient profile from source ($l=1$) to drain ($l=L$)
\begin{equation}\label{eq:ness_n}
    n_l = 1+j_\mathrm{NESS} \left( 1+ (l-1)\frac{\gamma}{2}+ \frac{1}{2}\delta_{l,1} - \frac12\delta_{l,L}\right)
\end{equation}
with a slope given by the average current $j_\mathrm{NESS} \sim 1/L$.  

The full dynamics was analyzed in Ref.~\cite{eisler2011crossover}. Within the Lindbladian framework, large deviations in the statistics of the current were later analyzed in Refs.~\cite{znidaric2014large,carollo2018current}. It is also important to mention that quantum trajectories in this model were analyzed before in Ref.~\cite{carollo2017fluctuating}, which, however, considered a \emph{unitary unraveling} of the averaged Lindbladian evolution. 
Within this framework, and for $\gamma >0$ the entanglement negativity has been estimated in Ref.~\cite{alba2021spreading}. 

\section{Fermionic negativity of Gaussian states}
\label{sec:fermionic_neg}

In this section we detail the computation of the fermionic negativity for Gaussian fermionic states. We consider a density matrix $\rho_{AB}$ over a bipartite system $A\cup B$, characterized by the $2N\times 2N$ Majorana matrix
\begin{equation}
	M{j,k}=\frac{1}{2} \operatorname{tr}\left(\rho\left[\psi_{j}, \psi_{k}\right]\right)\,.
\end{equation}
To compute Eq.~\eqref{eq:defneg}, the starting point is the block decomposition over $A\cup B$
\begin{equation}
	M=
	\begin{pmatrix}
		M_{AA} & 	M_{AB}\\
		M_{BA} & 	M_{BB}
	\end{pmatrix}\,.
\end{equation}
From~\eqref{eq:rhoR1}, it follows that 
\begin{equation}
	M_{\pm}=
	\begin{pmatrix}
		-M_{AA} & 	\pm i M_{AB}\\
		\pm i  M_{BA} & 	M_{BB}
	\end{pmatrix}\,.
\end{equation}
are the covariance matrices associated with $\rho^{R_A}$ and $[\rho^{R_A}]^{\dagger}$. The product of these covariance matrices can be performed following Refs.~\cite{fagotti2010entanglement,eisert2018entanglement}. The resulting density matrix is Gaussian with covariance matrix
\begin{equation}
M_{\ast}=\openone -\left(\openone - M_{-}\right)\frac{1}{\openone+M_+M_-}\left(\openone -M_+\right)\,.
\end{equation}
It can shown that this matrix is antisymmetric and purely immaginary, and so Hermitian. Furthermore, the product density matrix has a normalization factor given by~\cite{fagotti2010entanglement} $\sqrt{{\rm det}\left[{1+M^2}/{2}\right]}$. 

Collecting all the terms, we arrive at the final result
\begin{widetext}
\begin{align}
\mathcal{E}(\rho)
&=\sum_{j}\ln\left[\left(\frac{1+\xi_j}{2}\right)^{1/2}+\left(\frac{1-\xi_j}{2}\right)^{1/2}\right]+\frac{1}{2}\sum_{j}\ln\left[\left(\frac{1+\zeta^2_j}{2}\right)\right]\,.~\label{eq:finalneg}
\end{align}
Here we denoted by $\{\xi_j,-\xi_j\}$ and $\{\zeta_j,-\zeta_j\}$ the eigenvalues of $M_\ast$ and $M$, respectively (in both cases they come in pairs of opposite sign, because they are symmetric). So in the above sums only a single element in each pair must be included.
\end{widetext}
 
The formula Eq.~\eqref{eq:finalneg} holds for any quadratic fermionic system~\cite{alba2022logarithmic}. For the specific instance considered in the main text, and for the initial condition $\langle c^\dagger_a c_b^\dagger \rangle= 0 $, the computation of the negativity can be simplified. Defining the matrix $G_{m,n}=2C_{m,n} - \delta_{m,n}$, and given the bipartition $A\cup B$, we have
\begin{equation}
    G = \begin{pmatrix}
    G_{AA} & G_{AB} \\
    G_{BA} & G_{BB} 
\end{pmatrix}.
\end{equation}
Then, in a similar fashion to the Majorana case, one can compute
\begin{equation}
    G_\pm = \begin{pmatrix}
    G_{AA} & \pm i G_{AB} \\
    \pm i G_{BA} & -G_{BB} 
\end{pmatrix},
\end{equation}
and the matrix 
\begin{equation}
    G_\ast = \frac{1}{2}\left[\openone - (\openone + G_+ G_-)^{-1}(G_++G_-)\right].
\end{equation}
The final expression for the negativity for the restricted correlation is given by 
\begin{equation}
    \mathcal{E}(\rho) = \sum_{j} \left(\ln[\sqrt{\mu_j}+\sqrt{1-\mu_j}] + \frac{1}{2}\ln[1-2\lambda_j + 2\lambda_j^2]\right),
\end{equation}
where $\mu_j$ are the eigenvalues of $G_\ast$ and $\lambda_j$ are the eigenvalues of $C$. 

\section{Late-time scaling of the purity}
\label{sec:asymptotic_purity}

In this section, we provide an heuristic argument to justify that, for $\mu=1$, the average of the late-time R\'enyi-$2$ entropy is $O(1)$ as $L\to\infty$. First, following Ref.~\cite{fidkowski2021how}, we introduce the quantity
\begin{align}
	S_{\rm proxy}(C)&=2\ln(2){\rm tr}\left\{ \openone-\left[ C^2+(\openone-C)^2\right]  \right\}\,.
\end{align}
It is not difficult to show that, for a Gaussian state,
\begin{align}
	\frac{1}{2\ln(2)}S_{\rm proxy}(C)\leq S^{(2)}\leq S_{\rm proxy}(C)\,.
\end{align}
Therefore, $S_{\rm proxy}(C)$ and the R\'enyi-$2$ entropy $S^{(2)}$ have the same scaling in $L$, so that $S_{\rm proxy}$ can be considered a proxy for $S^{(2)}$. 

Next, we consider a simplified dynamics which we expect to display the same qualitative behavior of the SME~\eqref{eq:noise}. Namely, we focus on a discrete, rather than continuous, model where the single time-step consists in three parts: (i) the application of a quantum channel~\cite{nielsen2002quantum} acting at the boundary sites, implementing extraction/injection of particles; (ii) a finite-depth quantum-circuit Gaussian dynamics; (iii) a round of random measurements of the local density. For simplicity, we take the measurements to be projective, although our conclusions also hold if they are weak. They are performed at each site with a finite probability $p\in [0,1]$. In the following, we do not need to specify the quantum-circuit gates, which can be obtained, for instance, by a Trotterization of the Hamiltonian in Eq.~\eqref{eq:noise}. Finally, for the right quantum channel we choose a Gaussian operation implementing extraction of particles,
\begin{equation}\label{eq:channel}
\rho_S\mapsto {\rm tr}_A \{ \exp[-iH] (|0\rangle\langle 0|_{A} \otimes \rho_S ) \exp{iH} \}\,,
\end{equation}
where $H=(\pi/2)(c^\dagger_A c_L+ c^\dagger_L c_A)$. Here $c_A$, $c^\dagger_A$ act on an ancillary degree of freedom initialized in the vacuum, $\ket{0}$, and eventually they are traced over. Note that the symbol $\otimes$ in \eqref{eq:channel} denotes graded tensor product. Analogously, we can define a Gaussian operation implementing injection of particles at the left boundary.

Let us consider the stationary state for $\mu=1$. We want to estimate the variation of $S_{\rm proxy}$ due to the action of the right quantum channel, denoted by $\Delta S^{R}$. First, we recall that the average density at the right boundary site is $n_L=O(1/ L)$, cf. Eq.~\eqref{eq:ness_n}. Therefore, denoting by $C$ the covariance matrix along a typical trajectory, we have $C_{L,L}\sim O(1/ L)$ and $C_{j,L}, C_{L,j}\sim O(1/ \sqrt{L})$, for $j\neq L$. It follows that $\Delta S^R\sim 1/L$\,. The same holds for the left quantum channel, so that we can estimate the variation of $S_{\rm proxy}$ due to the action of the boundary channels as
\begin{equation}\label{eq:one}
\Delta S^{c}= c/L\,,
\end{equation}
for some constant $c$. Next, we would like to estimate the variation of $S_{\rm proxy}$ after the discrete steps (ii) and (iii) described above, which we denote by $\Delta S^m$. On average, the measurements decrease $S_{\rm proxy}$ since they have the tendency of purifying the system. Suppose we replace the finite-depth circuit with a random Gaussian unitary acting on the whole system. In this case, based on the analysis of Ref.~\cite{fidkowski2021how}, and using that we have on average $p L$ measurements, we would obtain the estimate $|\Delta S^{m}|\sim (S_{\rm proxy})^2/L$. For our quantum-circuit model, we expect $|\Delta S^{m}|$ to be larger than this (measurements purify more), because the unitary dynamics is less scrambling and so less effective in protecting quantum information from the measurements. Therefore, we obtain the lower bound
\begin{equation}\label{eq:two}
|\Delta S^{m}|\geq  c^\prime \frac{(S_{\rm proxy})^2}{L}\,.
\end{equation}
Now, after a single application of steps (i), (ii) and (iii), the total change of $S_{\rm proxy}$ is $\Delta S_{\rm proxy}=  \Delta S^{c}-| \Delta S^{m}|$. In the stationary regime, we must have $\Delta S_{\rm proxy}=0$. Putting together~\eqref{eq:one} and~\eqref{eq:two}, this readily implies 
\begin{equation}
S^2_{\rm proxy} \leq (c/c^\prime) \sim O(1)\,.
\end{equation}
Namely, even in the weak monitoring limit, the late-time R\'enyi-$2$ entropy does not grow with the system size, consistently with our numerical results. 

\begin{figure}[t!]
    \centering
    \includegraphics[width=\columnwidth]{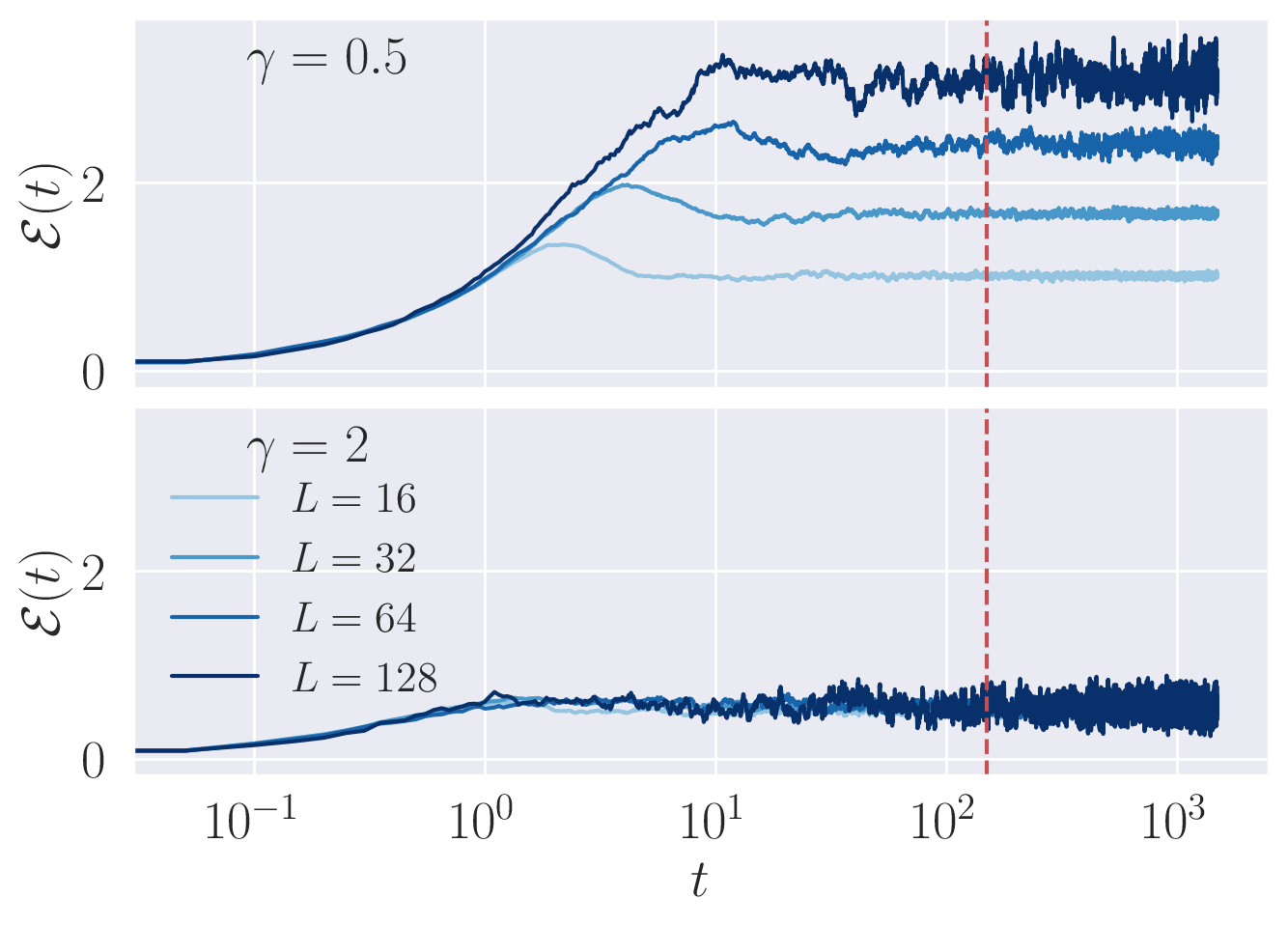}
    \caption{\label{fig:saturation} \textit{Choice of saturation time. ---} We consider $\Gamma=1$, and show which saturation time (red line) is chosen for $\gamma=0.5,\ 2$ and $L=16\div 128$. }
\end{figure}

\begin{figure}[t!]
    \centering
    \includegraphics[width=\columnwidth]{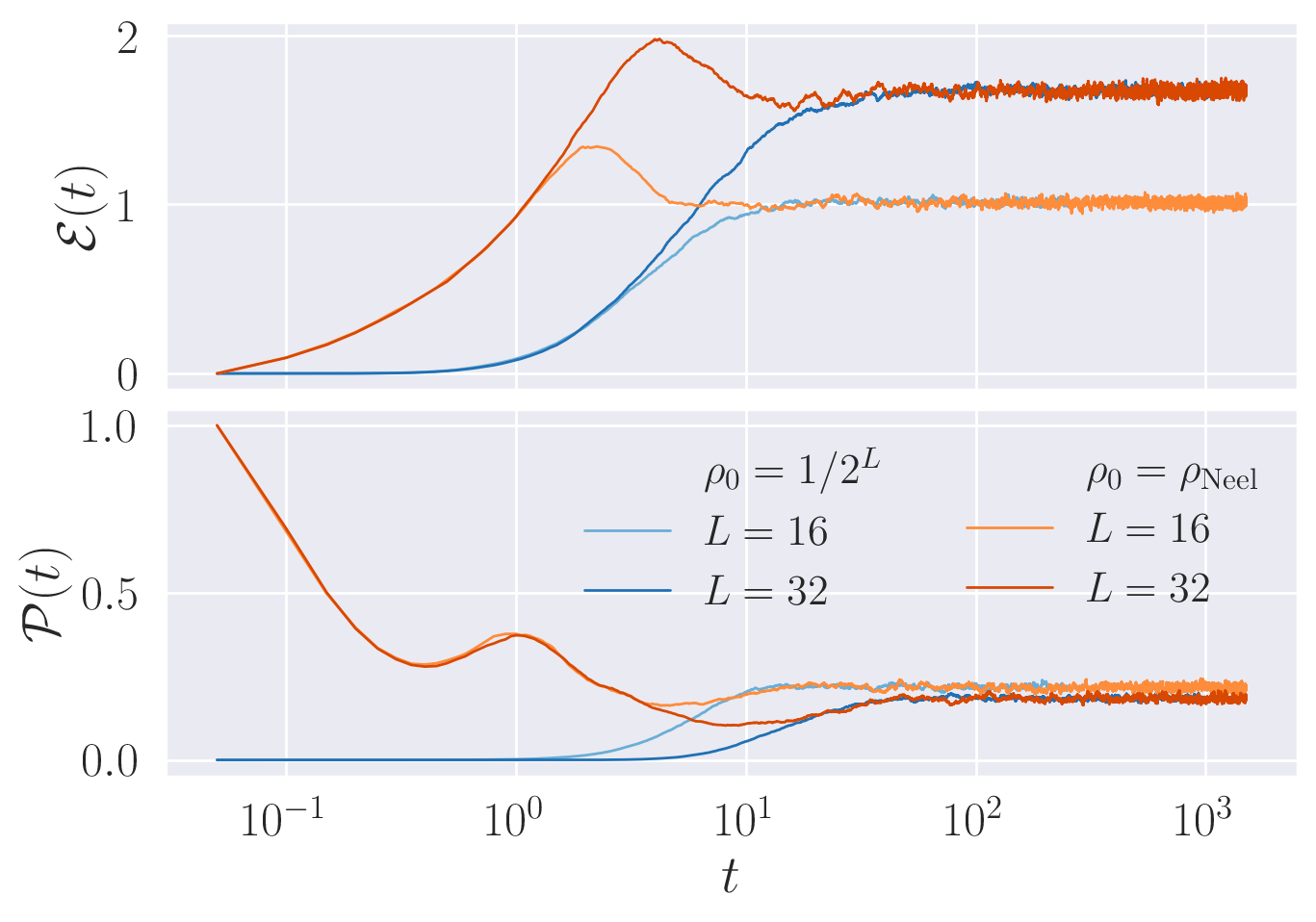}
    \caption{\label{fig:initialcond} \textit{Robustness against different initial conditions. ---} As an example, here we consider $\gamma=0.5$ and various system sizes with the infinite temperature initial state (blue) and the initial state~\eqref{eq:neel_state} (orange). As expected, the dynamics show a different transient regime, but the same stationary state.}
\end{figure}

\section{Numerical implementation and additional numerical benchmarks}
\label{sec:additional_checks}
In this section, we briefly summarize the numerical implementation and give further numerical results. 

We have used a Runge-Kutta algorithm of fourth order to solve the combined evolution consisting of boundary-driving and Hamiltonian dynamics and implemented the mapping Eq.~\eqref{eq:finalff} for the noise contribution. For the time-evolution of the quantum trajectories, we considered the average over $\mathcal{N}=200\div 1000$ trajectories. Furthermore, we have evidence that a self-averaging property occurs at late time: hence we also average the stationary state values over the last $T=5000\div 20000$ time-steps in the stationary regime. Throughout this paper, we have chosen $dt=0.05$, but we tested, but not shown here, that the protocol gives the same average results for $dt=0.01$. Another test we have performed is that the results are qualitatively robust against varying $\Gamma$, which have overall set to $\Gamma=1$ in this paper.

In Fig.~\ref{fig:saturation} we illustrate the choice of the stationary time $t_\mathrm{stat}$ for various $\gamma$ and various system sizes. Typically, due to the diffusive nature of the average state, the saturation time scales ad $O(L^2/\gamma)$, which combined with the $O(L^3)$ simulation cost of each time step, results in $O(L^5)$ cost.

\textit{Independence from the initial conditions.---}
Initial conditions affect the transient dynamics, but results in the same stationary state. As an example, in Fig.~\ref{fig:initialcond} we consider two different initial conditions, the infinite temperature state $\rho = 1/2^L$ and the N\'eel state~\eqref{eq:neel_state}. focusing on $\gamma=0.5$, $\Gamma=1$. Our numerics show the conditional average of the negativity and of the purity saturates to the same stationary value. 
We have also checked the independence from the initial conditions by taking random product states and different values of $\gamma$ (not shown here).

\bibliography{entanglement}
\bibliographystyle{apsrev4-1}

\end{document}